\documentclass[a4paper,10pt]{revtex4}
\usepackage{graphicx}  
\usepackage{amsmath}   
\usepackage{amssymb}   
\usepackage{bm} 
\usepackage{dcolumn}
\usepackage{color}
\usepackage{mathrsfs}
\usepackage{amsfonts}
\usepackage{varioref}
\usepackage{mathrsfs}
\usepackage{graphicx}
\usepackage{latexsym}
\usepackage{amsmath}
\usepackage{amssymb}
\usepackage{textcomp}
\usepackage{amsbsy}
\usepackage{graphics}
\usepackage{epstopdf}
\usepackage{color}
\usepackage[caption=false]{subfig}

\RequirePackage[colorlinks,citecolor=blue,urlcolor=magenta,linkcolor=blue]{hyperref}
\input epsf

\begin{document}

\tolerance=5000

\title{Ekpyrotic bounce driven by Kalb-Ramond field}

\author{Tanmoy~Paul$^{1,2}$\,\thanks{pul.tnmy9@gmail.com},
Soumitra~SenGupta$^{3}$\,\thanks{tpssg@iacs.res.in}} \affiliation{
$^{1)}$ National Institute of Technology Jamshedpur, Department of Physics, Jamshedpur - 831 014, India.\\
$^{2)}$ Labaratory for Theoretical Cosmology, International Centre of Gravity and Cosmos,
Tomsk State University of Control Systems and Radioelectronics (TUSUR), 634050 Tomsk, Russia\\
$^{3)}$ School of Physical Sciences, Indian Association for the Cultivation of Science, Kolkata-700032, India}


\tolerance=5000

\begin{abstract}
We propose an ekpyrotic bounce scenario driven by a second rank antisymmetric Kalb-Ramond field, where the universe initially contracts 
through an ekpyrotic stage having a non-singular bouncing like behaviour, and consequently, it smoothly transits to an expanding phase. 
In particular, the KR field has an interaction with a scalaron field (coming from higher curvature d.o.f) 
by a linear coupling. The interaction energy density between the KR and the scalaron proves to grow faster than $a^{-6}$ (with $a$ being the scale factor
of the universe) during the contraction phase -- which has negligible effects at the distant past, however as the universe continues to contract, 
this interaction energy density gradually grows and plays a significant role in violating the null energy condition or to trigger a non-singular bounce 
at a minimum value of the scale factor. 
The existence of the ekpyrotic phase justifies the resolution of the BKL instability, where the anisotropic energy density gets diluted compared to that 
of the bouncing agent. The bounce being symmetric and ekpyrotic, the energy density of the bouncing agent rapidly decreases after the bounce during the 
expanding phase of the universe, and consequently the standard Big-Bang cosmology gets recovered. 
In regard to the perturbation analysis, we find that the comoving Hubble radius diverges at the distant past, which 
leads to the generation era of the primordial perturbation modes at the deep contracting phase far away from the bounce. In effect, the curvature 
perturbation gets a blue tilted spectrum over the large scale modes -- not consistent with the Planck data. To circumvent this problem, we 
propose an extended scenario where the ekpyrotic phase is preceded by a 
quasi-matter dominated pre-ekpyrotic phase, and re-investigate the perturbation power spectrum. As a result, the 
primordial curvature perturbation, at scales that cross the horizon during the pre-ekpyrotic stage, turns out to be nearly scale invariant that is indeed 
consistent with the recent Planck data. The hallmark of the work is that although the Kalb-Ramond field has negligible footprints at present universe,
it has considerable impacts during early stage of the universe, which in turn makes the universe's evolution non-singular. 

\end{abstract}

\maketitle

\section{Introduction}
We are presently in an era wherein we have observational constraints on a variety of cosmological parameters 
describing the early and the late universe. These include parameters such as the scalar spectral index, the tensor to scalar 
ratio, running of index which describe the early universe, as well as the late-time equation of state (EoS) and Om(z) regarding the dark energy era. 
On the other hand, modern cosmology is still challenged by the question: Did the universe really start its expansion from a Big-Bang 
singularity or did the universe undergo a bounce? Inflation is an appealing cosmological scenario that can resolve the horizon and flatness problems, and 
most importantly, it predicts a nearly scale invariant perturbation power spectrum that is consistent with the observational data 
\cite{Guth:1980zm,Linde:2005ht,Langlois:2004de,Riotto:2002yw,Baumann:2009ds}. However, extrapolating 
backward in time, inflation hinges with a curvature singularity known as the Big-Bang singularity. Bouncing cosmology is an alternative of inflation, which 
is able to predict an observationally compatible perturbation power spectrum 
\cite{Brandenberger:2012zb,Brandenberger:2016vhg,Battefeld:2014uga,Novello:2008ra,Cai:2014bea,deHaro:2015wda,Raveendran:2018yyh,Raveendran:2018why,Martin:2001ue,Odintsov:2022unp,Buchbinder:2007ad,Brown:2004cs,Hackworth:2004xb,Peter:2002cn,Gasperini:2003pb,Creminelli:2004jg,Lehners:2015mra,Cai:2014xxa,Cai:2012va,Zhu:2021whu,Cai:2014zga,Avelino:2012ue,Barrow:2004ad,Haro:2015zda,
Elizalde:2014uba,Odintsov:2020fxb,Banerjee:2020uil}. 
However bouncing cosmology has an extra advantage that it leads to 
singularity free evolution of the universe. Actually in the context of bounce, the universe initially contracts and then bounces off from a minimum value of the 
scale factor, and consequently it smoothly enters to an expanding phase without encountering any curvature singularity.

Among various bounce models proposed so far, matter bounce scenario (MBS) gained a lot of attention as it produces a scale invariant power spectrum 
and also leads to a matter dominated universe at late time 
\cite{deHaro:2015wda,Cai:2008qw,Finelli:2001sr,Qiu:2010ch,deHaro:2012xj,Elizalde:2020zcb,Nojiri:2019lqw,WilsonEwing:2012pu}.
Despite the above mentioned successes, the MBS is riddled with some problems, like - (1) the tensor to scalar ratio 
in the scalar-tensor MBS becomes too large to be consistent with the observational data \cite{Akrami:2018odb}, (2) during the contracting phase of the MBS, 
the anisotropic energy density grows faster compared to that of the bouncing agent and hence the model suffers from the BKL instability \cite{new1}, 
(3) as the universe at late time is dominated by pressureless matter, the MBS does not predict the dark energy era of the universe -- 
not consistent with the supernovae observations \cite{Perlmutter:1996ds,Perlmutter:1998np,Riess:1998cb}. 
Successful attempts have been made to resolve these issues in modified theories of gravity 
\cite{Elizalde:2020zcb,Nojiri:2019lqw,Raveendran:2018why,Raveendran:2018yyh,Odintsov:2020zct,Odintsov:2021yva}. 
For example, some of our authors proposed a smooth unification from a bounce to the dark energy era 
with an intermediate deceleration era in higher curvature theories of gravity both in the case of symmetric and asymmetric bounce scenarios 
\cite{Odintsov:2020zct,Odintsov:2021yva}, 
where the generation era of primordial perturbation modes become different depending on the symmetric / asymmetric character of the bounce. However 
such unified bounce scenarios still suffer from the BKL instability. Thus, recently an unified scenario from an ekpyrotic bounce to the dark energy era 
has been proposed in \cite{Nojiri:2022xdo} where the anisotropic energy density gets diluted due to the existence of the ekpyrotic phase of contraction.

On a different side, a surprising feature of the present universe that despite having the signatures of rank two symmetric 
tensor field in the form of gravity, it carries no noticeable footprints of rank two antisymmetric tensor field, generally known as Kalb-Ramond field 
\cite{Kalb:1974yc}. 
Apart from the massless representation of the Lorentz group, the KR field also arise as closed string mode \cite{Duff}, and are of considerable interest, 
in the 
context of String theory. Recently the possible roles of Kalb-Ramond field (coupled to Bianchi-I geometry) in the Lorentz symmetry violation 
has been studied in \cite{Maluf:2021eyu}. In the arena of higher dimensional braneworld scenario or in higher curvature gravity theory, 
it has been shown that the KR 
coupling (with other normal matter fields) gets highly suppressed over the usual gravity-matter coupling, which explains why the KR field has negligible 
footprints at the present universe \cite{Mukhopadhyaya:2002jn,Das:2010xx,Das:2018jey}. 
However the KR field shows its considerable effects during the early stage of the universe \cite{Elizalde:2018rmz,Elizalde:2018now,Nair:2021flg,
Aashish:2018lhv,Aashish:2020mlw}. 
In the context of 
inflationary scenario, the presence of KR field slows down the acceleration of the universe and also modifies the inflationary observable parameters 
\cite{Elizalde:2018rmz,Elizalde:2018now}. 
For example the presence of the KR field makes the $F(R) = R+R^3$ inflationary model viable in respect to the Planck data, unlike to the case of 
$F(R) = R+R^3$ theory without the KR field which does not predict a good inflation at all \cite{Elizalde:2018rmz}. 

Motivated by the fact that the KR field shows considerable effects during the early inflationary universe 
(despite its negligible footprints at the late stage), here we explore its possible roles in driving a $non-singular~bounce$. In particular, we address 
an ekpyrotic bounce driven by the second rank antisymmetric Kalb-Ramond field in F(R) gravity theory. With a suitable conformal transformtion 
of the metric, the F(R) frame can be mapped to a scalar-tensor theory, where the KR field gets coupled with the scalaron field (coming from higher 
curvature d.o.f) by a simple linear coupling. Such interaction between the KR and the scalaron field proves to be useful in violating 
the null energy condition and to trigger a non-singular bounce. In regard to the perturbation analysis, we examine the curvature power spectrum for 
two different scenario depending on the initial conditions: (1) in the first scenario, the universe initially 
undergoes through an ekpyrotic phase of contraction and consequently the large scales of primordial 
perturbation modes cross the horizon during the ekpyrotic stage, while, (2) in the second scenario, the ekpyrotic phase is preceded by 
a quasi-matter dominated pre-ekpyrotic stage, and thus the large scale modes (on which we are interested) cross the horizon during the pre-ekpyrotic phase. 
The existence of the pre-ekpyrotic stage seems to be useful in getting a nearly scale invariant curvature perturbation spectrum over the large 
scale modes. The detailed qualitative features are discussed.

The following notations are used in the paper: $t$ is the cosmic time, $\eta$ represents the conformal time defined by $d\eta = dt/a(t)$ with $a(t)$ being 
the scale factor of the universe. An overdot denotes the derivative with respect to cosmic time, while a prime symbolizes $\frac{d}{d\eta}$.

\section{The model}
We start with a second rank antisymmetric Kalb-Ramond (KR) field in F(R) gravity, and the action is,
\begin{eqnarray}
 S = \int d^4x\sqrt{-g}\left[\left(\frac{1}{2\kappa^2}\right)F(R) - \frac{1}{12}H_{\mu\nu\alpha}H_{\rho\beta\delta}g^{\mu\rho}g^{\nu\beta}g^{\alpha\delta}
 \right]~,
 \label{jordan-action}
\end{eqnarray}
$\frac{1}{2\kappa^2}=M_\mathrm{Pl}^2$ ($M_\mathrm{Pl}$ being the Planck mass). 
$H_{\mu\nu\alpha}$ is the field strength tensor of KR field, defined by 
$H_{\mu\nu\alpha} = \partial_{[\mu}B_{\nu\alpha]}$. It may be noticed that $H_{\mu\nu\alpha}$ is invariant 
under the KR gauge transformation: $B_{\mu\nu}\rightarrow B_{\mu\nu} + \partial_{[\mu}\omega_{\nu]}$ that makes the 
action invariant under such transformation.\\
The above action can be mapped into the Einstein frame by using the following conformal transformation,
\begin{eqnarray}
 g_{\mu\nu} \longrightarrow \widetilde{g}_{\mu\nu} = \sqrt{\frac{2}{3}}\kappa\Phi(x^{\mu})~g_{\mu\nu}~,
 \label{conformal transformation}
\end{eqnarray}
with $\Phi$ being the conformal factor and related to the spacetime curvature as $\Phi(R) = \frac{1}{\kappa}\sqrt{\frac{3}{2}}F'(R)$. 
Consequently the action in Eq.(\ref{jordan-action}) can be expressed as a scalar-tensor theory \cite{Nojiri:2010wj},
\begin{eqnarray}
 S = \int d^4x\sqrt{-\widetilde{g}}\left[\frac{\widetilde{R}}{2\kappa^2} - \frac{3}{4\kappa^2}\left(\frac{1}{\Phi^2}\right)
 \widetilde{g}^{\mu\nu}\partial_{\mu}\Phi\partial_{\nu}\Phi - V(\Phi) - \frac{1}{12}\left(\sqrt{\frac{2}{3}}\kappa\Phi\right)H_{\mu\nu\rho}H^{\mu\nu\rho}
 \right]~.
 \label{Einstein-action}
\end{eqnarray} 
where $\widetilde{R}$ is the Ricci scalar formed by $\widetilde{g}_{\mu\nu}$ and 
$H^{\mu\nu\rho} = H_{\alpha\beta\delta}\widetilde{g}^{\mu\alpha}\widetilde{g}^{\nu\beta}\widetilde{g}^{\rho\delta}$. Clearly 
the scalar field arises from the higher curvature d.o.f, and the KR field gets coupled with the scalar field. The scalar field 
potential depends on the form of F(R), in particular,
\begin{eqnarray}
 V(\Phi) = \frac{1}{2\kappa^2}\left[\frac{RF'(R) - F(R)}{F'(R)^2}\right]~.
\end{eqnarray}
First we determine the energy-momentum tensor for the scalar field and the KR field,
\begin{eqnarray}
 T_\mathrm{\mu\nu}[\Phi] = \frac{3}{2\kappa^2}\left(\frac{1}{\Phi^2}\right)\partial_\mathrm{\mu}\Phi\partial_\mathrm{\nu}\Phi 
 - \widetilde{g}_\mathrm{\mu\nu}\left\{\frac{3}{4\kappa^2}\left(\frac{1}{\Phi^2}\right)\widetilde{g}^\mathrm{\alpha\beta}
 \partial_\mathrm{\alpha}\Phi\partial_\mathrm{\beta}\Phi + V(\Phi)\right\}~~,\nonumber\\
 T_\mathrm{\mu\nu}[B] = \frac{1}{6}\sqrt{\frac{2}{3}}~\kappa\Phi\left\{3\widetilde{g}_\mathrm{\nu\rho}H_\mathrm{\alpha\beta\mu}
 H^\mathrm{\alpha\beta\rho} - \frac{1}{2}\widetilde{g}_\mathrm{\mu\nu}H_\mathrm{\alpha\beta\gamma}H^\mathrm{\alpha\beta\gamma}\right\}~.
 \label{EM tensor}
\end{eqnarray}
Here we are interested in the cosmological evolution of the KR field. For that purpose, we assume the ansatz of a flat FRW metric:
\begin{eqnarray}
 ds^2 = -dt^2 + a^2(t)\delta_\mathrm{ij}dx^\mathrm{i}dx^\mathrm{j}
 \label{FRW metric}
\end{eqnarray}
where $t$ and $a(t)$ are the cosmic time and the scale factor of the universe respectively. 
Here we would like to emphasize that $H_{\mu\nu\lambda}$ has four 
independent components in four dimensional spacetime due to its antisymmetric nature, and they can be expressed as,
\begin{eqnarray}
 H_\mathrm{012} = h_\mathrm{1}~~~,~~~H_\mathrm{013} = h_\mathrm{2}~~~,~~~H_\mathrm{023} = h_\mathrm{3}~~~,~~~H_\mathrm{123} = h_\mathrm{4}~,
 \label{independent components-KR}
\end{eqnarray}
and for $H^{\mu\nu\lambda}$,
\begin{eqnarray}
 H^\mathrm{012} = h^\mathrm{1}~~~,~~~H^\mathrm{013} = h^\mathrm{2}~~~,~~~H^\mathrm{023} = h^\mathrm{3}~~~,~~~H^\mathrm{123} = h^\mathrm{4}~.
\end{eqnarray}
Eq.(\ref{independent components-KR}) along with the expressions for the energy-momentum tensor and the FLRW metric lead to the 
off-diagonal Friedmann equations as follows,
\begin{eqnarray}
 h_\mathrm{1}h^\mathrm{2} = h_\mathrm{1}h^\mathrm{3} = h_\mathrm{2}h^\mathrm{3} = h_\mathrm{1}h^\mathrm{4} 
 = h_\mathrm{2}h^\mathrm{4} = h_\mathrm{3}h^\mathrm{4} = 0~,
 \label{off-diagonal-E-equations}
\end{eqnarray}
where the fields are considered homogeneous. Solving the above set of equations,
\begin{eqnarray}
 h_\mathrm{1} = h_\mathrm{2} = h_\mathrm{3} = 0~~~~~~~~\mathrm{and}~~~~~~~h_\mathrm{4} \neq 0~.
 \label{solution-off-diagonal-E-equation}
\end{eqnarray}
Here it deserves mentioning that the FRW metric ansatz of Eq.(\ref{FRW metric}) along with the antisymmetric nature of $B_{\mu\nu}$ (or the gauge invariance of $B_{\mu\nu}$) allow the above solution of KR field strength tensor. Therefore we may argue that, similar to the metric ansatz, this solution of KR field tensor with $h_1 = h_2 = h_3 = 0$ and $h_4 \neq 0$ obeys the rotational symmetry of the spatial part. Actually the presence of KR field with all $h_i$ ($i = 1,2,3,4$) are non-zero breaks the rotational symmetry of the spatial part of spacetime, however the said symmetry can be achieved in the case where the first three components of the KR field tensor vanish and $h_4 \neq 0$. We may note that in each of the first three components of the KR field tensor, one of the spatial indices is absent (for example, $h_1 = H_{012}$ where the 'z' index is absent); however on contrary, $h_4 = H_{123}$ contains all the spatial indices. Therefore the solution in Eq.(\ref{solution-off-diagonal-E-equation}) with only $h_4$ being non-zero indicates that the three spatial directions are on equal footing, which results to the rotational symmetry of the system. Moreover, being only one non-zero component of the KR field tensor, we may argue that the KR field has one degree of freedom which can be equivalently thought as of a scalar field; and a scalar field allows the rotational symmetry of the spatial part in a FRW spacetime.\\

Using the solution of Eq.(\ref{off-diagonal-E-equations}), one easily obtains the total energy density and pressure for the matter fields ($\Phi$, $B_{\mu\nu}$) as,
\begin{eqnarray}
\rho&=&\frac{3}{4\kappa^2}\left(\frac{\dot{\Phi}}{\Phi}\right)^2 + V(\Phi) + \frac{1}{2}\sqrt{\frac{2}{3}}\kappa\Phi~h_\mathrm{4}h^\mathrm{4}~~,\nonumber\\
p&=&\frac{3}{4\kappa^2}\left(\frac{\dot{\Phi}}{\Phi}\right)^2 - V(\Phi) + \frac{1}{2}\sqrt{\frac{2}{3}}\kappa\Phi~h_\mathrm{4}h^\mathrm{4}
\label{energy density-pressure}
\end{eqnarray}
respectively. Consequently the diagonal Friedmann equations takes the following form,
\begin{eqnarray}
 H^2&=&\frac{\kappa^2}{3}
 \left[\frac{3}{4\kappa^2}\left(\frac{\dot{\Phi}}{\Phi}\right)^2 + V(\Phi) + \frac{1}{2}\sqrt{\frac{2}{3}}\kappa\Phi~h_\mathrm{4}h^\mathrm{4}\right]~,
 \label{diagonal-E-equation-0}\\
 2\dot{H}&+&3H^2 
 + \kappa^2\left[\frac{3}{4\kappa^2}\left(\frac{\dot{\Phi}}{\Phi}\right)^2 - V(\Phi) 
 + \frac{1}{2}\sqrt{\frac{2}{3}}\kappa\Phi~h_\mathrm{4}h^\mathrm{4}\right] = 0~,
 \label{diagonal-E-equations}
\end{eqnarray}
where $H=\frac{\dot{a}}{a}$ is the Hubble parameter. 
Furthermore, the field equations for the scalar field ($\Phi$) and the KR field ($B_{\mu\nu}$) turn out to be,
\begin{eqnarray}
 \frac{\ddot{\Phi}}{\Phi} - \left(\frac{\dot{\Phi}}{\Phi}\right)^2 + 3H\left(\frac{\dot{\Phi}}{\Phi}\right) 
 + \left(\frac{2\kappa^2}{3}\right)\Phi~V'(\Phi) + \left(\frac{\sqrt{2}}{3\sqrt{3}}\right)\kappa^3\Phi~h_\mathrm{4}h^\mathrm{4} = 0
 \label{scalar field equation}
\end{eqnarray}
and
\begin{eqnarray}
 \frac{1}{a^3}\partial_{\mu}\left(a^3\Phi H^{\mu\nu\lambda}\right) = 0
 \label{KR field equation-1}
\end{eqnarray}
respectively. From Eq.(\ref{KR field equation-1}), 
we can argue that the non-zero component of $H_{\mu\nu\alpha}$ 
(i.e $H_{123}=h_4$) depends on $t$ only (see Appendix Sec.[\ref{sec-appendix-1}] for the derivation), 
which is also expected from the gravitational field equations. 
Differentiating both sides (with respect to $t$) of the temporal component of Einstein equation, one gets
\begin{eqnarray}
 6H\dot{H} = \kappa^2\left[\frac{3}{2\kappa^2}\left\{\frac{\dot{\Phi}\ddot{\Phi}}{\Phi^2} - \frac{\dot{\Phi}^3}{\Phi^3}\right\} 
 + V'(\Phi)\dot{\Phi} + \frac{\kappa}{2}\sqrt{\frac{2}{3}}\frac{d}{dt}\left(\Phi h_4h^{4}\right)\right]~.
\end{eqnarray}
Substituting the scalar field equation of motion in the above expression, we obtain the following cosmic evolution for $h_4h^4$ 
(see Appendix in Sec.[\ref{sec-appendix-new}] for the details):
\begin{eqnarray}
 \frac{d}{dt}\left(h_\mathrm{4}h^\mathrm{4}\right) + 6Hh_\mathrm{4}h^\mathrm{4} = 0~.
 \label{KR field equation-2}
\end{eqnarray}
Thus as a whole, the field equations in the present context are given by Eq.(\ref{diagonal-E-equation-0}), Eq.(\ref{diagonal-E-equations}), 
Eq.(\ref{scalar field equation}) and Eq.(\ref{KR field equation-2}) respectively. However all these four equations are not independent, in particular, one 
of them can be derived from the others. Thus we take the following three equations as independent field equations:
\begin{eqnarray}
 3H^2 - \kappa^2\left[\frac{3}{4\kappa^2}\left(\frac{\dot{\Phi}}{\Phi}\right)^2 + V(\Phi) 
 + \frac{1}{2}\sqrt{\frac{2}{3}}\kappa\Phi~h_\mathrm{4}h^\mathrm{4}\right]&=&0~~,\label{independent-1}\\
 2\dot{H} + \kappa^2\left[\frac{3}{2\kappa^2}\left(\frac{\dot{\Phi}}{\Phi}\right)^2 + 
 \sqrt{\frac{2}{3}}~\kappa\Phi h_4h^4\right]&=&0~~,\label{independent-2}\\
 \frac{d}{dt}\left(h_\mathrm{4}h^\mathrm{4}\right) + 6Hh_\mathrm{4}h^\mathrm{4}&=&0~~.
 \label{independent-3}
\end{eqnarray}
The above equations need to be simultaneously solved to get the corresponding evolutions of the Kalb-Ramond field, scalar field and the Hubble parameter 
of the universe. It may be observed that Eq.(\ref{independent-3}) provides the Kalb-Ramond field in terms of the Hubble parameter, thus it can 
immediately solved to obtain the KR field in terms of scale factor, as follows,
\begin{eqnarray}
 h_\mathrm{4}h^\mathrm{4} = h_\mathrm{0}/a^\mathrm{6}~,
 \label{solution-KR field}
\end{eqnarray}
with $h_0$ being an integration constant which is taken to be positive to ensure a real valued solution for $h^4(t)$. 
We now turn to the other two equations which are the coupled equations between the scalar field and the Hubble parameter, and the important 
point to be mentioned that Eq.(\ref{independent-2}) does not contain the $V(\Phi)$. Therefore our strategy is as follows: (1) we first solve 
Eq.(\ref{independent-2}) to get $H = H(a)$ by considering an ansatz for the scalar field evolution in terms of the scale factor, and then (2) 
by plugging all the field solutions into Eq.(\ref{independent-1}), we will reconstruct the form of $V(\Phi)$ such that the solutions we obtain 
simultaneously satisfy the independent field equations. 

We consider an ansatz for the scalar field solution in terms of the scale factor as,
\begin{eqnarray}
 \Phi(a) = -\frac{1}{\kappa}\sqrt{\frac{3}{2}}\left(\frac{1}{a^\mathrm{n}}\right)~,
 \label{ansatz-scalar field}
\end{eqnarray}
with $n > 0$. Eq.(\ref{ansatz-scalar field}) indicates that the scalar field acquires negative values during the cosmological evolution of the universe, 
which actually proves to be useful to get a non-singular bounce. As we will show later, the presence of such scalar field 
will not harm the stability of the primordial perturbation. With the above solutions for $\Phi(t)$ and $h_4(t)$, the temporal and spatial 
components of Einstein equation (Eq.(\ref{diagonal-E-equations})) become,
\begin{eqnarray}
 H^2 = \frac{\kappa^2}{3\left(1 - n^2/4\right)}\left[V(\Phi(t)) - \frac{h_\mathrm{0}}{2a^\mathrm{6+n}}\right]
 \label{temporal E-equation}
\end{eqnarray}
and
\begin{eqnarray}
 \left(aH\right)\frac{dH}{da} + \frac{3}{4}n^2H^2 - \frac{\kappa^2h_\mathrm{0}}{2a^\mathrm{6+n}} = 0
 \label{spatial E-equation}
\end{eqnarray}
respectively. Eq.(\ref{spatial E-equation}) provides a two branch solution of $H = H(a)$:
\begin{eqnarray}
 H(a) = \pm a^\mathrm{-3n^2/4}\left\{C_\mathrm{1} + \frac{2\kappa^2h_\mathrm{0}a^\mathrm{\left(3n^2-2n-12\right)/2}}{\left(3n^2-2n-12\right)}\right\}^{1/2}~,
 \label{solution-Hubble parameter}
\end{eqnarray}
where $C_1$ is an integration constant. Eq.(\ref{solution-Hubble parameter}) clearly demonstrates that the Hubble parameter 
acquires a negative branch as well as a positive branch solution, 
which can be identified with a contracting and an expanding universe respectively. 
This leads to a natural possibility of bounce in the present context. 
Moreover the evolution of $H(a)$ becomes different depending on whether $3n^2-2n-12 < 0$ or 
$3n^2-2n-12 > 0$. However both the cases will be proved to lead a non-singular bounce irrespective of the values of 
$n > 0$. Moreover, the bounce scenario will be shown to be free from the anisotropic instability for suitable regime of $n$. 
First we discuss the case when $3n^2-2n-12 < 0$ and its consequences, 
while the other case having a similar analysis is described in the Appendix Sec.[\ref{sec-appendix-2}]. 
We consider $3n^2 - 2n - 12 = -2q$, with $q > 0$, and consequently, the solution of $H(a)$ in Eq.(\ref{solution-Hubble parameter}) can be 
expressed as,
\begin{eqnarray}
 H(a) = \pm \sqrt{\frac{\kappa^2h_\mathrm{0}}{q}}~a^\mathrm{-3n^2/4}\left\{\frac{1}{a_\mathrm{0}^q} - \frac{1}{a^q}\right\}^{1/2}
 \label{solution-Hubble parameter-case-new1}
\end{eqnarray}
where the integration constant $C_1$ is replaced by $a_0$ as $C_\mathrm{1} = \kappa^2h_\mathrm{0}/\left(qa_\mathrm{0}^q\right)$. 
Such an evolution of the Hubble parameter allows a non-singular universe that bounces at $a = a_0$, we will demonstrate it in the next section. 
Plugging back the expression of Eq.(\ref{solution-Hubble parameter-case-new1}) 
into Eq.(\ref{temporal E-equation}), we reconstruct the scalar field potential in terms of scale factor as follows,
\begin{eqnarray}
 V(\Phi(a)) = h_0~a^{-3n^2/2}\left[\frac{3}{q}\left(1 - \frac{n^2}{4}\right)\frac{1}{a_0^q} 
 + \frac{1}{a^{q}}\left\{\frac{1}{2} - \frac{3}{q}\left(1 - \frac{n^2}{4}\right)\right\}\right]~~.
\end{eqnarray}
Finally using Eq.(\ref{ansatz-scalar field}) into the above expression to get the potential in terms of scalar field as,
\begin{eqnarray}
 V(\Phi) = h_0\left(-\sigma\right)^{3n/2}\left[\frac{3}{q}\left(1 - \frac{n^2}{4}\right)\frac{1}{a_0^q} 
 + \left(-\sigma\right)^{q/n}\left\{\frac{1}{2} - \frac{3}{q}\left(1 - \frac{n^2}{4}\right)\right\}\right]~,
 \label{potential-1}
\end{eqnarray}
where $\sigma = \sqrt{\frac{2}{3}}\kappa\Phi$ and recall that the scalar field $\Phi$ acquires negative values (see Eq.\ref{ansatz-scalar field}). 
Thus as a whole, the solutions of the KR field, the scalar field and the Hubble parameter (in terms of the scale factor) are given by,
\begin{eqnarray}
 h_4h^4&=&h_0/a^6~~,\nonumber\\
 \Phi(a)&=&-\frac{1}{\kappa}\sqrt{\frac{3}{2}}\left(\frac{1}{a^\mathrm{n}}\right)~~,\nonumber\\
 H(a)&=&\pm a^\mathrm{-3n^2/4}\left\{C_\mathrm{1} + \frac{2\kappa^2h_\mathrm{0}a^\mathrm{\left(3n^2-2n-12\right)/2}}{\left(3n^2-2n-12\right)}
 \right\}^{1/2}~~,
\label{new-solution}
\end{eqnarray}
respectively, and moreover, the $V(\Phi)$ is reconstructed in Eq.(\ref{potential-1}). Clearly with this $V(\Phi)$, the above set of solutions 
simultaneously satisfy the independent field equations given from Eq.(\ref{independent-1}) to Eq.(\ref{independent-3}). Here it may be observed that during the reconstruction procedure, $V(\Phi)$ seems to depend on 
$h_0$ which appears as an integration constant (with mass dimension [+4]) in Eq.(\ref{solution-KR field}) -- this demonstrates that 
$h_0$ actually represents the parameter of the scalar field potential. 

\begin{figure}[!h]
\begin{center}
 \centering
 \includegraphics[width=3.0in,height=2.5in]{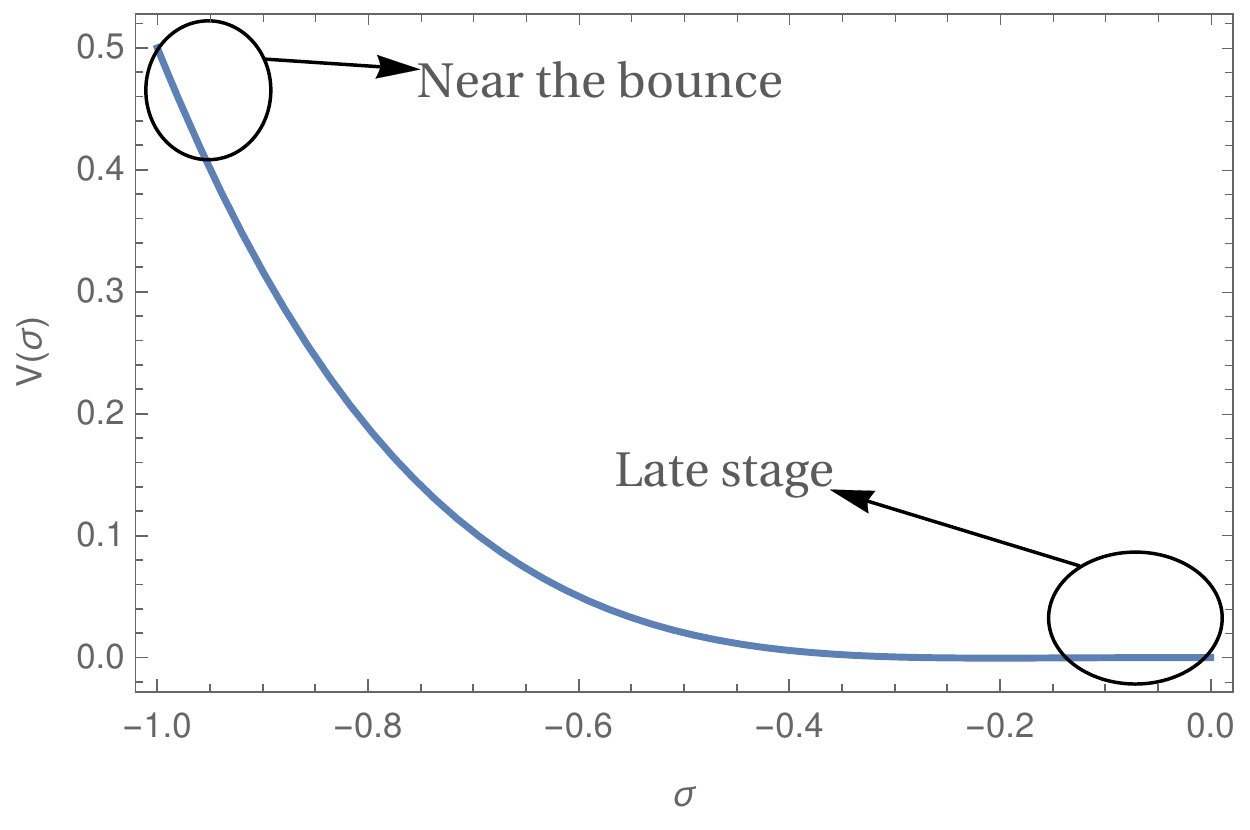}
 \includegraphics[width=3.0in,height=2.5in]{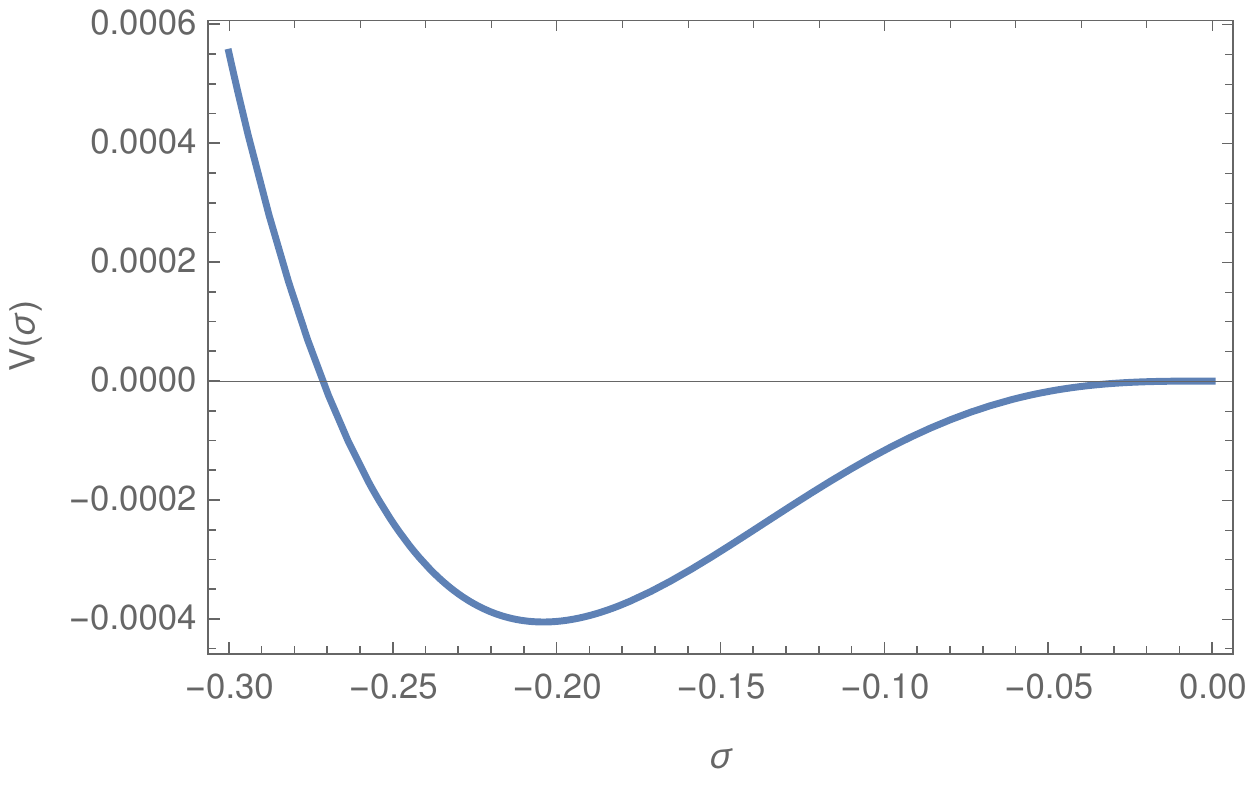}
 \caption{$V(\Phi)$ (in the unit of $h_0$) vs. $\sigma$ from Eq.(\ref{potential-1}). Here we take $n=2.2$ and $a_0=1$. Due to $a_0=1$, 
 $\sigma = -1$ at the bounce and $\sigma \rightarrow 0^{-}$ far away at both sides of the bounce -- thus the range of $\sigma$ is given by 
 $-1 \leq \sigma < 0^{-}$. The right part of the figure demonstrates the zoomed-in version of $V(\Phi)$ near the stable minimum.}
 \label{plot-potential-1}
\end{center}
\end{figure}

Coming back to Eq.(\ref{potential-1}), due to the ekpyrotic condition $n>2$, 
the $V(\Phi)$ becomes negative during the late contraction or late expansion era of the 
universe. Moreover the scalar field potential has a stable minimum at,
\begin{eqnarray}
 \Phi = \frac{1}{a_0^n\kappa}\sqrt{\frac{3}{2}}\left[\frac{9n\left(\frac{n^2}{4}-1\right)}{\left(n+6\right)}\right]^{n/q}\nonumber~.
\end{eqnarray}
We give a plot of $V(\Phi)$ (with respect to $\sqrt{\frac{2}{3}}\kappa\Phi = \sigma$) in Fig.[\ref{plot-potential-1}] where we take $a_0=1$, for which, 
$\sigma = -1$ at the bounce and $\sigma \rightarrow 0^{-}$ far away at both sides of the bounce (see Eq.(\ref{ansatz-scalar field})) -- 
thus the range of $\sigma$ is given by $-1 \leq \sigma < 0^{-}$ for $a_0=1$. The right part of 
the Fig.[\ref{plot-potential-1}] is the zoomed-in version of $V(\Phi)$ near the stable minimum of the scalar field. It is evident that 
the $V(\Phi)$ becomes negative near the stable point -- this is due to the ekpyrotic condition $n>2$, as mentioned earlier.\\

During the deep contracting phase, the scalar field starts to evolve from $\sigma \rightarrow 0^{-}$ (recall $\sigma \propto -1/a^{n}$), 
and it rolls along the potential as the universe evolves. Consequently at the bouncing point when the Hubble parameter vanishes, the scalar field 
reaches to the top most value of the potential (as indicated in the Fig.[\ref{plot-potential-1}]) where the velocity of the scalaron 
(i.e $\dot{\sigma}$) turns out to be zero. Being $\dot{\sigma} = 0$, the scalar field traces back its path along the potential 
during the expanding phase of the universe and finally $\sigma$ goes to $0^{-}$ at the late expanding stage.\\

Here we would like to mention that in order to solve the field equations, we consider an ansatz between the scalar field and the scale factor as in Eq.(\ref{ansatz-scalar field}). This is because the form of F(R) leads to one functional degree of freedom which actually allows the relation like Eq.(\ref{ansatz-scalar field}). Based on this argument, we now demonstrate the functional form of F(R) corresponding to Eq.(\ref{ansatz-scalar field}). Let us concentrate during the expanding phase of the universe when the Hubble parameter is given by Eq.(\ref{solution-Hubble parameter-case-new1}) with the positive sign. Then the Ricci scalar in the scalar-tensor theory turns out to be (symbolized by $\widetilde{R}$, see Eq.(\ref{Einstein-action})):
\begin{eqnarray}
 \widetilde{R}(a) = \left(\frac{3\kappa^2h_0}{2q}\right)\left(8 - 3n^2\right)a_0^{-q}~a^{-3n^2/2}\left\{1 - \left(\frac{3n^2 + 2q - 8}{3n^2 - 8}\right)\left(\frac{a_0}{a}\right)^q\right\}~~.
 \label{rev-1}
\end{eqnarray}
Owing to the conformal transformation as of Eq.(\ref{conformal transformation}), the Ricci scalar in the scalar-tensor theory is connected to that of in the conformally connected F(R) theory by,
\begin{eqnarray}
 R = \sigma \left[\widetilde{R} + \frac{9}{2}\left(\frac{\dot{\sigma}}{\sigma}\right)^2 - 9H\left(\frac{\dot{\sigma}}{\sigma}\right) - 3\left(\frac{\ddot{\sigma}}{\sigma}\right)\right]~~,
 \label{rev-2}
\end{eqnarray}
where recall that $R$ is the Ricci scalar in the F(R) frame. By using Eq.(\ref{ansatz-scalar field}), we get $\frac{\dot{\sigma}}{\sigma} = nH$ which, along with the solution of the Hubble parameter from Eq.(\ref{solution-Hubble parameter-case-new1}), lead to the above expression as follows:
\begin{eqnarray}
 R = \sigma^{(3n+2)/2}\left(\frac{3\kappa^2h_0}{4q}\right)\left(3n - 4\right)\left(n^2 - 4\right)\left\{1 - \left(\frac{3n^2 + 2n + 2q - 8}{\left(3n - 4\right)\left(n + 2\right)}\right)\left(\frac{a_0}{a}\right)^q\right\}~~.
 \label{rev-3}
\end{eqnarray}
Due to $\sigma = F'(R)$, we can write Eq.(\ref{rev-3}) as,
\begin{eqnarray}
 R \propto \left(\frac{dF}{dR}\right)^{(3n+2)/2}\left\{1 - a_0^q\left(\frac{3n^2 + 2n + 2q - 8}{\left(3n - 4\right)\left(n + 2\right)}\right)\left(\frac{dF}{dR}\right)^{q/n}\right\}~~.
 \label{rev-4}
\end{eqnarray}
Eq.(\ref{rev-4}) is a differential equation for F(R), on solving which, one will get the form of F(R). However as we note that Eq.(\ref{rev-4}) is highly non-linear and may not be solved analytically. However during the expanding universe when $a \gg a_0$ happens (that surely happens after some e-fold number from the time of bounce), the second term inside the curly braket of Eq.(\ref{rev-3}) can be ignored and thus Eq.(\ref{rev-3}) can be safely written as,
\begin{eqnarray}
 R \propto \sigma^{(3n+2)/2}~~.
 \label{rev-5}
\end{eqnarray}
Then, due to $\sigma = F'(R)$, the differential equation is given by: $F'(R) \propto R^{2/(3n+2)}$, on solving which, we obtain
\begin{eqnarray}
 F(R) \propto R^m~~~;~~~~~~~~~\mathrm{with}~~~~~~~~~m = \frac{3n + 4}{3n + 2}~~.
 \label{rev-6}
 \end{eqnarray}
 Thus as a whole, the form of F(R) in the present context obeys the non-linear Eq.(\ref{rev-4}) which has an analytic solution as $F(R) \propto R^m$ during the expanding phase of the universe away from the bounce.

 Here it may be mentioned that the same form of F(R) as of Eq.(\ref{rev-6}) can also be obtained from the procedure discussed in \cite{Nojiri:2010wj} that how a scalar-tensor theory with a non-zero scalar potential can be equivalently mapped to a F(R) theory by a conformal transformation of the spacetime metric. According to \cite{Nojiri:2010wj}, the form of F(R) corresponding to a scalar-tensor theory having a scalar potential $V(\sigma)$ is given by,
 \begin{eqnarray}
  F(R) = \left(\frac{R}{2\kappa^2}\right)\sigma - V(\sigma)\sigma^2~~,
  \label{rev-7}
 \end{eqnarray}
where the scalar field can be obtained in terms of $R$, i.e $\sigma = \sigma(R)$, by solving the following algebraic equation:
\begin{eqnarray}
 R = 2\kappa^2 \sigma\left[2V(\sigma) + \sigma~\frac{dV}{d\sigma}\right]~~.
 \label{rev-8}
\end{eqnarray}
For the scalar potential in the present context of Eq.(\ref{potential-1}), the above expression immediately leads to,
\begin{eqnarray}
 R \propto \sigma^{(3n+2)/2}~~~;~~~~~~~~~~\mathrm{or}~~~~~~~~~\sigma(R) \propto R^{2/(3n+2)}~~,
 \label{rev-9}
\end{eqnarray}
where we retain the term up-to the leading order in $\sigma$, which is viable during the expanding phase of the universe away from the bounce. By plugging the above $\sigma = \sigma(R)$ into Eq.(\ref{rev-7}) and with a little bit of simplification yields the following form of F(R) during the expanding stage of the universe as,
\begin{eqnarray}
 F(R) \propto R^{\frac{(3n+4)}{(3n+2)}}~~.
 \label{rev-10}
\end{eqnarray}
This form of F(R) clearly matches with that of in Eq.(\ref{rev-6}) obtained from the previous procedure.

\section{Realization of a non-singular bounce with the condition: $3n^2 - 2n - 12 < 0$}\label{sec-background}
\begin{figure}[!h]
\begin{center}
 \centering
 \includegraphics[width=3.5in,height=2.5in]{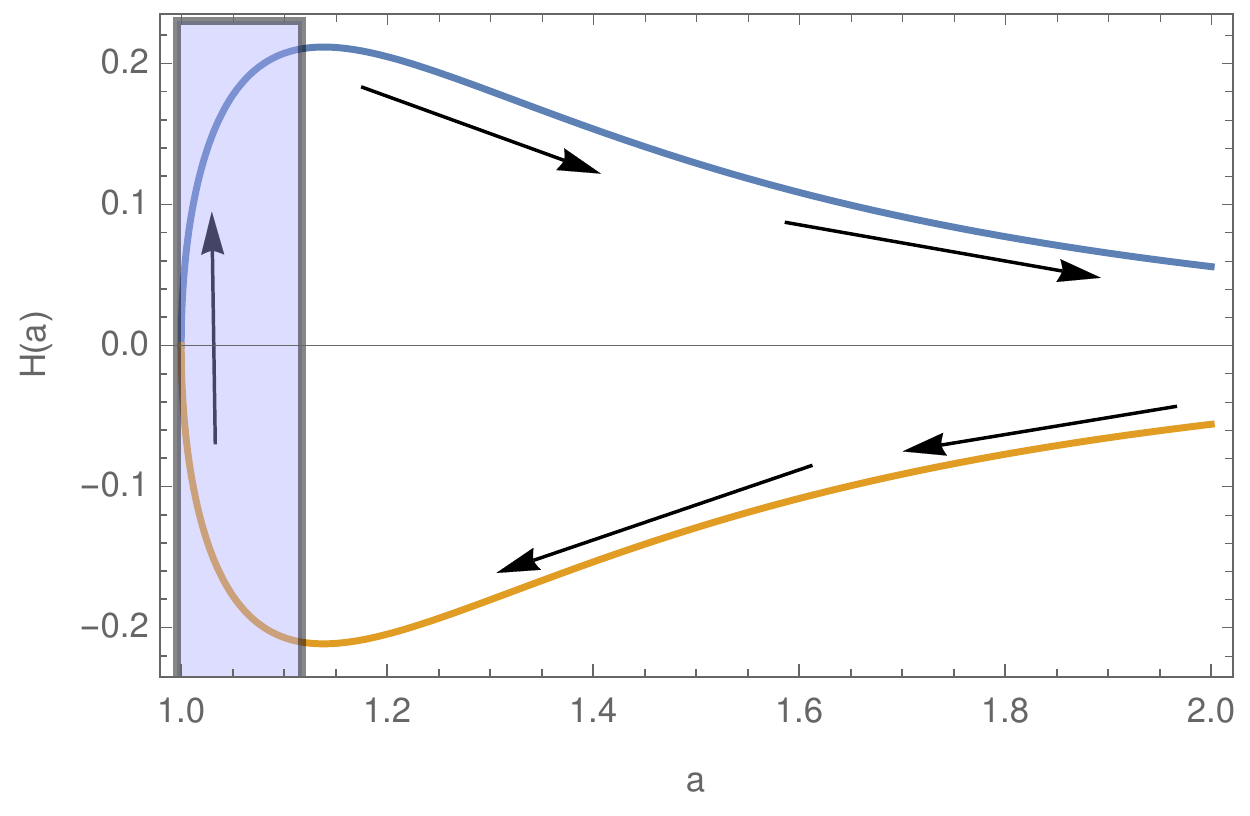}
 \caption{$H(a)$ vs. $a$ from Eq.(\ref{solution-Hubble parameter-case-1}) with $a_0 = 1$. The yellow and blue curves represent 
 the negative and the positive branch of $H(a)$ respectively.}
 \label{plot-Hubble-case-1}
\end{center}
\end{figure}

For the case $3n^2 - 2n - 12 = -2q$ ( with $q > 0$ ), recall the evolution of the Hubble parameter from 
Eq.(\ref{solution-Hubble parameter-case-new1}) given by,
\begin{eqnarray}
 H(a) = \pm \sqrt{\frac{\kappa^2h_\mathrm{0}}{q}}~a^\mathrm{-3n^2/4}\left\{\frac{1}{a_\mathrm{0}^q} - \frac{1}{a^q}\right\}^{1/2}~~.
 \label{solution-Hubble parameter-case-1}
\end{eqnarray}
The above expression of $H(a)$ satisfies the following two conditions at $a = a_0$,
\begin{eqnarray}
 H(a=a_0) = 0~~~~~~~~~~\mathrm{and}~~~~~~~~~~~\frac{dH}{dt}\bigg|_{a_0} = \left(aH\right)\frac{dH}{da}\bigg|_{a_0} = \frac{\kappa^2h_0}{2a_0^{6+n}}~,
 \label{bounce-confirm}
\end{eqnarray}
which clearly depicts that the universe experiences a non-singular bounce at $a=a_0$. For a better understanding, 
we give a plot of $H(a)$ (with respect to $a$) in Fig.[\ref{plot-Hubble-case-1}] using Eq.(\ref{solution-Hubble parameter-case-1}), 
where the yellow and blue curves represent the negative and positive branch of $H(a)$ respectively. The figure clearly demonstrates that 
the Hubble parameter starts to evolve from the negative branch (i.e $H(a) < 0$) and results in to a contracting 
phase of the universe. Going forward in time through the negative branch, the universe continues to contract and finally at $a = a_0$, 
$H(a)$ becomes zero and $\dot{H}(a)$ gets a positive value. Consequently, the Hubble parameter smoothly enters to the positive branch (i.e $H(a)>0$) that 
leads to an expanding phase of the universe. In particular, the Hubble parameter flows along the directed arrow shown in the 
Fig.[\ref{plot-Hubble-case-1}] from the distant past to the distant future. 
This ensures a non-singular bounce at $a = a_0$, which in turn provides a possible explanation for the removal of Big-Bang singularity.\\ 

Here it deserves mentioning that in absence of the KR field, the model is not able to 
predict a non-singular bounce of the universe. In particular, the Hubble parameter evolves as 
$H(a) \propto a^{-3n^2/4}$ when the KR field is absent, 
which does not lead to a bouncing scenario at all. However the presence of the KR field 
considerably modifies the evolution of $H(a)$ and plays a significant role in getting a non-singular bounce. Actually the interaction 
energy density between the KR and the scalaron field varies as $\sim a^{-(6+n)}$ that has negligible effects at the distant past 
compared to that of the scalaron field self energy density. However as the universe continues to contract, such interaction 
energy density gradually grows and plays an important role to trigger a non-singular bouncing universe. Due to this Kalb-Ramond interaction 
with the scalaron field, the null energy condition gets violated near $a=a_0$, which makes the bounce possible. 
Following, we examine the energy condition during the universe's evolution in the present context. Using Eq.(\ref{energy density-pressure}), we get,
\begin{eqnarray}
 \rho + p = h_0~a^{-3n^2/2}\left\{\left(\frac{3n^2}{2q}\right)\frac{1}{a_0^q} - \left(1 + \frac{3n^2}{2q}\right)\frac{1}{a^q}\right\}~.
 \label{energy condition-case-1}
\end{eqnarray}
The above expression reveals that $\rho+p = -h_0/a^{(6+n)} < 0$ at $a=a_0$, i.e the energy condition seems to be violated 
at the bounce, as expected. Furthermore, during the late contracting or late expanding phase (when $a \gg a_0$), $\rho+p$ becomes positive, i.e there is 
no energy condition violation during the late stage of the universe. In particular, the null energy condition in the present context 
remains violated during $a < a_c$ where,
\begin{eqnarray}
 a_c =  a_0\left[\frac{2(6+n)}{3n^2}\right]^{1/q}\nonumber~~.
\end{eqnarray}
Thereby the bouncing epoch can be identified as $a < a_c$. Interestingly $H'(a)$ vanishes at $a=a_c$. Thus the epoch of bounce can be represented by the 
shaded region in the Fig.[\ref{plot-Hubble-case-1}]. 

Before proceeding further, here we would like to mention the consequences of the transformation (\ref{conformal transformation}) where 
the scalar field acquires negative values during the universe's evolution (see Eq.(\ref{ansatz-scalar field})). 
Let us recall the F(R) action along with the matter field action in Eq.(\ref{jordan-action}) given by,
\begin{eqnarray}
 S = \int d^4x\sqrt{-g}\left[\left(\frac{1}{2\kappa^2}\right)F(R) - \frac{1}{12}H_{\mu\nu\alpha}H_{\rho\beta\delta}g^{\mu\rho}g^{\nu\beta}g^{\alpha\delta}\right]~.
 \label{jordan-action-new}
\end{eqnarray}
For a transformation like, 
\begin{eqnarray}
 g_{\mu\nu} \rightarrow \widetilde{g}_{\mu\nu} = -g_{\mu\nu}~,
 \label{new-trans}
\end{eqnarray}
the gravity part of the above action, namely,
\begin{eqnarray}
 S_\mathrm{grav} = \int d^4x\sqrt{-g}\left[\left(\frac{1}{2\kappa^2}\right)F(R)\right]~.
 \label{jordan-action-new-1}
\end{eqnarray}
remains same. However due to the presence 
of odd number of  metric functions, the matter part of the action, i.e
\begin{eqnarray}
 S_\mathrm{mat} = \int d^4x\sqrt{-g}\left[- \frac{1}{12}H_{\mu\nu\alpha}H_{\rho\beta\delta}g^{\mu\rho}g^{\nu\beta}g^{\alpha\delta}\right]~,
 \label{jordan-action-new-2}
\end{eqnarray}
gets transformed by a negative sign. Such transformation of $S_\mathrm{mat}$ introduces a ghost matter field 
in the ``transformed tilde'' frame. This analogy can drawn in the present context where the metric transformation is given by 
Eq.(\ref{conformal transformation}) where $\Phi$ itself is negative. 
As a result, the Kalb-Ramond (KR) field in the the scalar-tensor theory emerges as a ghost field, given by action (\ref{Einstein-action}) where 
the kinetic term of the KR field becomes negative during the universe's evolution due to the coupling 
between the KR field and the scalar field of the scalar-tensor theory.

At this stage it deserves mentioning that the ghost behaviour of matter field and the consequent violation of  the null energy condition 
is a generic feature of having a non-singular bounce in the realm of classical effective theory (for instance, see \cite{Cai:2012va}). 
However it may be observed that an important estimate i.e. 
the ratio of the Kalb-Ramond to the scalaron field energy density here is given by,
\begin{eqnarray}
 \frac{\rho_{KR}}{\rho_{\Phi}} = \left[\frac{\frac{1}{2}\sqrt{\frac{2}{3}}\kappa\Phi~h_\mathrm{4}h^\mathrm{4}}
 {\frac{3}{4\kappa^2}\left(\frac{\dot{\Phi}}{\Phi}\right)^2 + V(\Phi)}\right]~~.\nonumber
\end{eqnarray}
Due to the fact that the scalar field acquires negative values (see Eq.(\ref{ansatz-scalar field})), the kinetic term of the KR field 
gets negative during the universe's evolution. However at the late contracting as well as at the 
late expanding phase of the universe, i.e away from the bounce, $\rho_{KR}$ goes as $\sim a^{-(n+6)}$ while $\rho_{\Phi} \sim a^{(q-n-6)}$. 
This demonstrates that the KR field energy density is heavily suppressed in comparison to the scalar field energy density, which in turn does not
violate the total energy condition  during the later phase of the evolution of the  universe. Therefore the universe, far away from the bounce, is not affected 
by the negative kinetic energy of the KR field. However around the bouncing regime, both the KR field and the scalar field energy density turn 
out to be comparable to each other, in particular, both the 
$\rho_{KR}$ and $\rho_{\Phi}$ evolve as $\sim a^{-(6+n)}$ near the bounce. 
Therefore the negative kinetic energy of the KR field affects the evolution of the universe, and consequently, 
violates the energy condition around the bouncing regime. Thus in regard to the effective total energy density ( in particular, 
$\rho_\mathrm{eff} = \rho_\mathrm{KR} + \rho_\mathrm{\Phi}$ ), 
we may argue that the ghost nature reflected in  the effective energy density 
is transient only  around the bounce.

\subsection*{Condition for $ekpyrotic$ character of the bounce}
During the deep contracting era, the Hubble parameter evolves as $H(a) \propto a^{-3n^2/4}$ (from Eq.(\ref{solution-Hubble parameter-case-1})), 
which immediately leads to the effective equation of state (EoS) parameter as,
\begin{eqnarray}
 \omega_\mathrm{eff} = -1 - \left(\frac{2a}{3H}\right)\frac{dH}{da} = -1 + \frac{n^2}{2}~.
 \label{eos-case-1}
\end{eqnarray}
This implies that the contracting phase of the universe is driven by super stiff matter, or equivalently $\omega_\mathrm{eff} > 1$, for $n >2$. 
Since $\omega_\mathrm{eff} > 1$, the energy density $\rho$ grows faster than $a^{-6}$ 
during ekpyrotic contraction. Such a behavior allows one to circumvent the difficulty posed by the rapid growth of 
anisotropies (which behave as $a^{-6}$) that proves to be a great drawback affecting many of the bouncing scenarios \cite{Belinskii,Levy:2016xcl}. 
Here we would like 
to mention that the bounce being ekpyrotic and symmetric, the energy density of the bouncing agent becomes proportional to 
$a^{-3n^2/2}$ after the bounce during the expanding phase, which rapidly decreases 
(faster than that of the pressureless matter and radiation due to the ekpyrotic condition $n>2$) 
as the universe expands, and consequently the standard Big-Bang 
cosmology of the universe is recovered. It is important to realize that a rank two KR field which has negligible footprints 
at present universe, seems to play the most important role during the early phase of the universe to trigger an ekpyrotic bounce.\\

Moreover the condition $3n^2-2n-12 < 0$ leads to $\frac{1}{3}\left(1-\sqrt{37}\right) < n < \frac{1}{3}\left(1+\sqrt{37}\right)$. Therefore in 
order to have an ekpyrotic character of the bounce, the parameter $n$ is constrained by,
\begin{eqnarray}
 2 < n < \frac{1}{3}\left(1 + \sqrt{37}\right)~.
 \label{constraint-1}
\end{eqnarray}
Using such parametric regime, we now analyze the curvature perturbation in the next section.\\

\section{Perturbation analysis}
Recall from Eq.(\ref{solution-Hubble parameter-case-1}) that the Hubble parameter at the distant contracting phase 
evolves by $H(a) \propto -a^{-3n^2/4}$. Therefore the comoving hubble radius (defined by $r_h = \left|\frac{1}{aH}\right|$) scales as 
$r_h \propto a^{(3n^2 - 4)/4}$ which, due to the ekpyrotic condition $n>2$, diverges to infinity at the deep contracting phase. 
This indicates that 
the primordial perturbation modes generate during the contracting phase far away from the bounce when all the perturbation modes lie within the 
Hubble horizon. Here we would like to mention that the generation of the perturbation in the contracting phase is consistent with the 
ekpyrotic character of the bounce, due to the fact that the effective EoS parameter during the ekpyrotic contraction is larger than unity. 
Such generation era of the primordial perturbation further ensures the resolution of the horizon problem in the bouncing scenario. 
Moreover, since the model involves two fields, apart from the curvature perturbation, 
isocurvature perturbation also arises. As mentioned earlier, the Kalb-Ramond to the scalaron field energy 
density during the late stage of the universe goes by $1/a^{q}$ which tends to zero. This 
results to a weak coupling between the curvature and the isocurvature perturbations during the late evolution of the universe. 
However the ratio between $\rho_{KR}$ and $\rho_{\Phi}$ becomes comparable at the bounce, which in turn leads to a considerable coupling 
between the curvature and the isocurvature perturbations at the bounce. In the present context, we are interested 
to examine the curvature perturbation during the contracting universe (away from the bounce), and thus, 
we can safely ignore the coupling between the curvature and the isocurvature perturbations and solely concentrate on the curvature perturbation.

The Mukhanov-Sasaki (MS) variable of the curvature perturbation follows the equation (in the Fourier space and conformal time coordinate):
\begin{eqnarray}
 v_k''(\eta) + \left(k^2 - \frac{z''}{z}\right)v_k(\eta) = 0~,
 \label{MS-equation-1}
\end{eqnarray}
where $\eta$ is the conformal time, $v_k(\eta)$ is the MS variable and a prime denotes $\frac{d}{d\eta}$. 
Moreover the function $z(\eta)$ is given by,
\begin{eqnarray}
 z(\eta) = \frac{a}{\kappa}\left(\frac{\dot{\Phi}}{\Phi H}\right)~.
 \label{z-1}
\end{eqnarray}
At this stage it deserves mentioning that the factor $z(\eta)$ actually appears in front of the 
kinetic term of the curvature perturbation variable 
in the second order scalar perturbed action over the FRW spacetime. Thus $z(\eta)$ not only determines the interaction of the 
curvature perturbation with the background 
FRW spacetime, but also controls the stability of the curvature perturbation. In particular, the curvature perturbation is stable if the condition 
$z^2(\eta) > 0$ is fulfilled during the universe's evolution. Eq.(\ref{z-1}) along with the field solutions obtained in Eq.(\ref{new-solution}) 
lead to the following form of $z^2(\eta)$ in the present context,
\begin{eqnarray}
 z^2(\eta) = \left(\frac{n^2}{\kappa^2}\right)a^2(\eta)~~.
 \label{new-z}
\end{eqnarray}
This demonstrates that $z^2$ is positive, which ensures the stability of the curvature perturbation during the universe's evolution. Moreover 
Eq.(\ref{MS-equation-1}) indicates that the speed of the scalar perturbation, i.e $c_s^2$, is unity in the present scenario -- this removes 
the possibility of the gradient instability of the curvature perturbation. 
Using the above form of $z(\eta)$, Eq.(\ref{MS-equation-1}) 
turns out to be,
\begin{eqnarray}
 v_k''(\eta) + \left(k^2 - \frac{a''}{a}\right)v_k(\eta) = 0~.
 \label{MS-equation-2}
\end{eqnarray}
As mentioned earlier, the perturbation modes generate and cross the horizon during the ekpyrotic contraction, and thus we intend to solve 
Eq.(\ref{MS-equation-2}) during the same where $H(a) \propto -a^{3n^2/4}$ or equivalently $a(\eta) \propto \left(-\eta\right)^{4/(3n^2-4)}$. 
Hence Eq.(\ref{MS-equation-2}) becomes,
\begin{eqnarray}
 v_k''(\eta) + \left(k^2 - \frac{4\left(8-3n^2\right)}{\left(3n^2-4\right)^2\eta^2}\right)v_k(\eta) = 0~,
 \label{MS-equation-3}
\end{eqnarray}
on solving which for $v_k(\eta)$, we get,
\begin{eqnarray}
v(k,\eta) = \frac{\sqrt{\pi|\eta|}}{2}\left[c_1(k)H_{\nu}^{(1)}(k|\eta|) + c_2(k)H_{\nu}^{(2)}(k|\eta|)\right]\, ,
\label{MS-solution}
\end{eqnarray}
with $\nu = \sqrt{\frac{4\left(8-3n^2\right)}{\left(3n^2-4\right)^2} + \frac{1}{4}}$. 
Moreover $c_1$, $c_2$ are integration constants, $H_{\nu}^{(1)}(k|\eta|)$ and $H_{\nu}^{(2)}(k|\eta|)$ are the Hermite functions (having order 
$\nu$) of first and second kind, respectively. 
The perturbation modes start from the Bunch-Davies vacuum state results to $\lim_{k|\eta| \gg 1}v(k,\eta) = \frac{1}{\sqrt{2k}}e^{-ik\eta}$ 
which is indeed consistent with the Eq.(\ref{MS-equation-3}) at $\eta \rightarrow -\infty$. 
Owing to the Bunch-Davies condition, the integration constants are given by $c_1 = 0$ and $c_2 = 1$, respectively. 
Consequently, the curvature power spectrum for $k$-th mode turns out to be,
\begin{eqnarray}
\mathcal{P}(k,\eta) = \frac{k^3}{2\pi^2} \left| \frac{v(k,\eta)}{z(\eta)} \right|^2 
= \frac{k^3}{2\pi^2} \left| \frac{\sqrt{\pi|\eta|}}{2z (\eta)}H_{\nu}^{(2)}(k|\eta|) \right|^2~,
\label{curvature-power-spectrum}
\end{eqnarray}
where in the second equality, we use the solution of $v(k,\eta)$. 
The horizon crossing condition for $k$-th mode is given by $k = |aH|$ which, due to $a(\eta) \propto \left(\-\eta\right)^{4/\left(3n^2-4\right)}$, 
can be simplified as,
\begin{eqnarray}
k\left| \eta_h\right| = \frac{4}{3n^2 - 4}~,
\label{hc-1}
\end{eqnarray}
where the suffix 'h' represents the horizon crossing instant of the $k$th mode. This demonstrates 
the sub-Hubble and super-Hubble regime of $k$-th mode in the present context as follows,
\begin{align}
k\left|\eta\right|>\frac{4}{3n^2 - 4}\, :& \, \mathrm{sub~Hubble~regime}\, ,\nonumber\\
k\left|\eta\right|<\frac{4}{3n^2 - 4}\, :& \, \mathrm{super~Hubble~regime}\, .
\label{hc-2}
\end{align}
Here we need to recall that $n>2$ from the requirement of an ekpyrotic phase of contraction, 
by which, the quantity $\frac{4}{3n^2 - 4}$ becomes less than unity. 
Thereby from Eq.(\ref{hc-2}), one may equivalently express the super-Hubble regime by $k\left|\eta\right| \ll 1$. 
As a result, the curvature power spectrum (from Eq.(\ref{curvature-power-spectrum})) in the super-Hubble scale is given by,
\begin{eqnarray}
\mathcal{P}(k,\eta) = \left[\left(\frac{1}{2\pi}\right)\frac{1}{z\left|\eta\right|}\frac{\Gamma(\nu)}{\Gamma(3/2)}\right]^2
\left(\frac{k|\eta|}{2}\right)^{3-2\nu}\, .
\label{curvature-power-spectrum-superhorizon}
\end{eqnarray}
Using the above expression, we can calculate the spectral tilt of the primordial curvature perturbation ($n_s$) defined by,
\begin{eqnarray}
 n_s = 1 + \frac{\partial\ln{\mathcal{P}}}{\partial\ln{k}}\bigg|_{h.c}~,\nonumber
\end{eqnarray}
and has a constraint according to the Planck 2018 data,
\begin{eqnarray}
 n_s = 0.9649 \pm 0.0042~.
 \label{Planck}
\end{eqnarray}
Due to Eq.(\ref{curvature-power-spectrum-superhorizon}), the $n_s$ in the present context comes with the following form,
\begin{eqnarray}
 n_s = \frac{9n^2 - 4}{3n^2 - 4}~.
 \label{spectral-tilt-1}
\end{eqnarray}
As demonstrated in Eq.(\ref{constraint-1}), the parameter $n$ is constrained by $2 < n < \frac{1}{3}\left(1 + \sqrt{37}\right)$, 
which immediately leads to the following range of the spectral tilt: $3.6 < n_s < 4$. This indicates a blue-tilted curvature power spectrum, 
and thus is not consistent with the Planck 2018 results.

\subsection*{A pre-ekpyrotic phase and a nearly scale invariant power spectrum}
\begin{figure}[!h]
\begin{center}
\centering
\includegraphics[width=3.5in,height=3.0in]{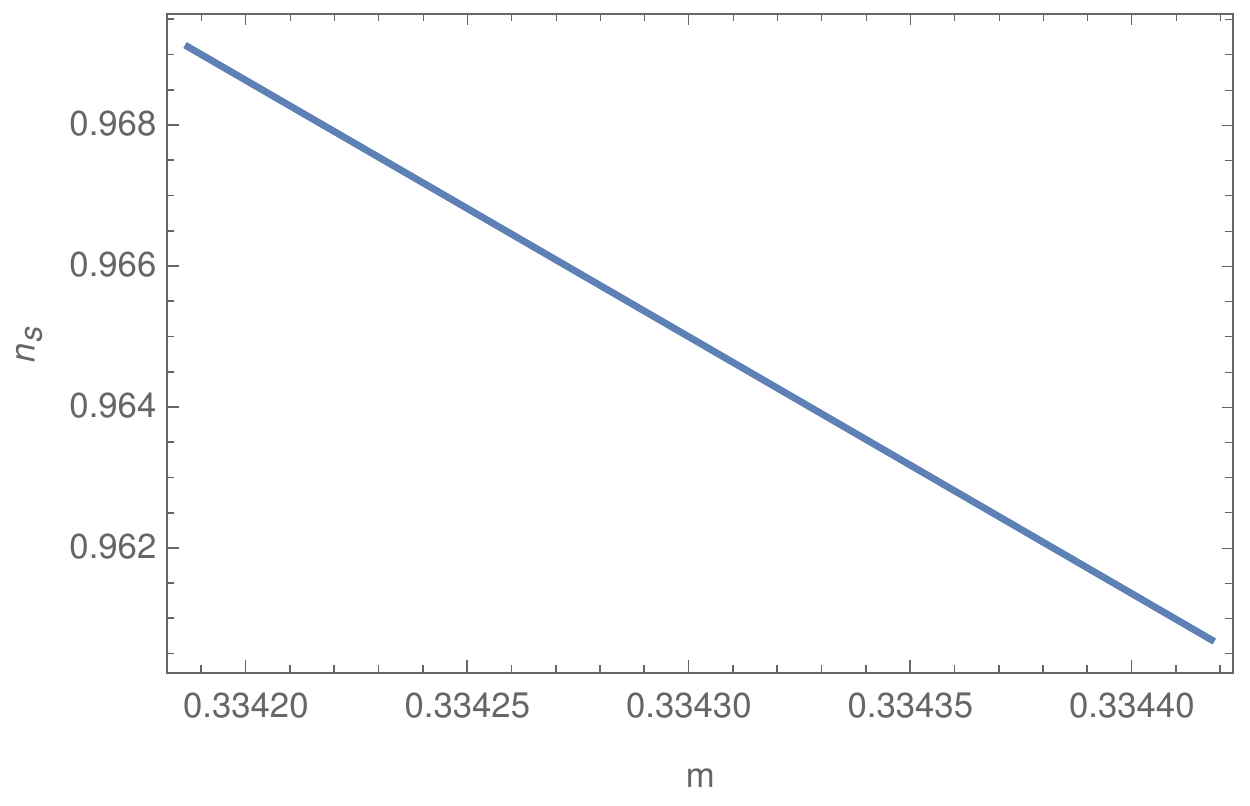}
\caption{$n_s$ vs. $m$ from Eq.(\ref{obs-1-pre})}
\label{plot-observable}
\end{center}
\end{figure}

We have demonstrated that the cosmological scenario, where the universe is dominated by a phase of ekpyrotic contraction before the bounce 
in the context of Kalb-Ramond coupled scalaron theory, leads to a blue tilted curvature power spectrum 
which is not consistent with the Planck results. 
Such an inconsistency arises because the large scale perturbation modes generate from an ekpyrotic vacuum fluctuations. 
This finding is in agreement with \cite{Cai:2012va,Cai:2014zga} where the authors studied an ekpyrotic bounce scenario from a different perspective 
without/with accounting loop quantum effects around the bounce, unlike to the present scenario where a second rank antisymmetric KR field gets coupled 
with a scalaron field. In order to get a scale invariant curvature power spectrum in the present context, 
we consider a quasi-matter dominated pre-ekpyrotic phase where the scale factor behaves as $a_p(t) \sim t^{2m}$ with $m < 1/2$. Such a quasi-matter 
dominated phase can be realized by introducing a perfect fluid having constant EoS parameter $\approx 0$, in which case, the energy density 
grows as $\approx a^{-3}$ during the contracting phase. Thereby after some time, the KR field energy density (that grows as $a^{-(6+n)}$ with
the universe's contraction, see Eq.(\ref{solution-KR field})) dominates over that of the perfect fluid 
and leads to an ekpyrotic phase of the universe. In particular, 
the pre-ekpyrotic scale factor is described by (\cite{Cai:2012va,Cai:2014zga}),
\begin{eqnarray}
a_p(t) = a_\mathrm{1}\left(\frac{\eta - \eta_0}{\eta_e - \eta_0}\right)^{2m/\left(1-2m\right)} \quad \mbox{with} \quad m<1/2\, ,
\label{pre-scale factor}
\end{eqnarray}
and re-investigate the curvature power spectrum. Here $\eta_e$ represents the conformal time when the transition from the pre-ekpyrotic to the 
ekpyrotic phase occurs, and $\eta_0$ is a fiducial time. Moreover the exponent $m < 1/2$ so that the comoving Hubble radius 
diverges at the distant past and the perturbation modes generate at the deep contracting phase within the sub-Hubble regime. 
As a whole, in this modified cosmological scenario, the scale factor of the universe is:
\begin{align}
a_p(t) = a_\mathrm{1}\left(\frac{\eta - \eta_0}{\eta_e - \eta_0}\right)^{2m/\left(1-2m\right)} \quad &\mbox{with} \quad m<1/2\, ,  
\quad \mathrm{for}~\left|\eta\right| \geq \left|\eta_e\right|\, ,\nonumber\\
a(t) = a_2\left(-\eta\right)^{4/\left(3n^2 - 4\right)} \quad &\mbox{with} \quad 2<n<\frac{1}{3}\left(1+\sqrt{37}\right) \, , 
\quad \mathrm{for}~\left|\eta\right| \leq \left|\eta_e\right|\, .
\label{full-scale-factor}
\end{align} 
The continuity of the scale factor as well as of the Hubble parameter at the transition time $\eta=\eta_e$ result to the following expressions, 
\begin{eqnarray}
a_1&=&a_2\left(-\eta_e\right)^{4/\left(3n^2 - 4\right)}~,\nonumber\\
\left(\frac{2m}{1-2m}\right)\frac{1}{\left(\eta_e - \eta_0\right)}&=&\left(\frac{4}{3n^2 - 4}\right)\frac{1}{\eta_e}~,
\label{continuity}
\end{eqnarray}
respectively. In effect of the pre-ekpyrotic phase of contraction, the large scale perturbation modes cross the horizon either during the pre-ekpyrotic 
or during the ekpyrotic stage depending on whether the transition time ($\eta_e$) is larger than the horizon crossing instant of the large scale modes 
($\eta_h$) or not. For a scale invariant power spectrum, here 
we consider the first case, i.e when the large scale modes cross the horizon during the pre-ekpyrotic phase. 
Thus the horizon crossing instant for $k$th mode is given by,
\begin{eqnarray}
\left|\eta_h\right| = \left(\frac{2m}{1-2m}\right)\frac{1}{k}\, .
\label{hc-1-pre-ekpyrotic}
\end{eqnarray}
Consequently the horizon crossing instant for the large scale modes, in particular $k = 0.002\mathrm{Mpc}^{-1}$, is estimated as,
\begin{eqnarray}
\left|\eta_h\right| \approx \left(\frac{2m}{1-2m}\right)\times13\,\mathrm{By}\, .
\label{hc-2-pre-ekpyrotic}
\end{eqnarray}
Thus one may argue that the transition from the pre-ekpyrotic to the ekpyrotic phase occurs at $\left|\eta_e\right| < 13\mathrm{By}$ so that 
the large scale modes cross the horizon during the pre-ekpyrotic era. Following 
the same procedure as of the previous section, we calculate the spectral tilt for the curvature perturbation in the modified 
scenario where the ekpyrotic phase is preceded by a period of a pre-ekpyrotic contraction:
\begin{eqnarray}
n_s = \frac{5- 14m}{1 - 2m}\, .
\label{obs-1-pre}
\end{eqnarray}
Clearly for $m = 1/3$ which describes a matter dominated epoch before the ekpyrotic phase, the spectral tilt becomes unity -- i.e an exact scale invariant 
power spectrum is predicted when the curvature perturbations over the large scale modes generate during a matter dominated pre-ekpyrotic era. 
However the observations according to the Planck data depict that the curvature power spectrum should not be exactly flat, but a has a slight red tilt 
as mentioned in Eq.(\ref{Planck}). For this purpose, we give a plot of $n_s$ with respect to $m$ in Fig.[\ref{plot-observable}]. The figure clearly 
demonstrates that the theoretical prediction of $n_s$ becomes consistent with the Planck 2018 data if the parameter $m$ lies 
within $0.3341 \lesssim m \lesssim 0.3344$. Therefore the spectral index for the primordial curvature perturbation, on scales that 
cross the horizon during the pre-ekpyrotic stage with $0.3341 \lesssim m \lesssim 0.3344$, is found to be consistent with the recent Planck 
observations.\\

Before concluding, here we would to like mention that in addition to the scalar type perturbation, primordial tensor perturbation is 
also generated in the deep contracting phase from the Bunch-Davies state. 
The recent Planck data puts an upper bound on the tensor perturbation amplitude, in particular on the 
tensor to scalar ratio as $< 0.064$. However the Mukhanov-Sasaki equation for the 
tensor perturbation in the present context becomes analogous compared to that of the scalar perturbation, and thus 
both type of perturbations evolve in a similar way. This makes the tensor to scalar ratio in the 
present bounce model too large to be consistent with the Planck observation. There are some ways to circumvent this problem, like - 
(1) by the curvaton mechanism considered in \cite{Cai:2011zx}, (2) by amplifying the curvature perturbation from the gradient instability of 
$c_\mathrm{s}^2$ (sound speed) changing sign during the bounce, (3) by introducing Gauss-Bonnet higher curvature terms in the action \cite{Nojiri:2022xdo} 
etc. This will be an interesting generalization of the present scenario by introducing such mechanisms that may 
reduce the tensor to scalar ratio. We leave this particular topic for future study.

\section{Conclusion}
We propose a bouncing scenario driven by a second rank antisymmetric Kalb-Ramond (KR) field in F(R) gravity, 
in which the universe initially contracts through an ekpyrotic phase having a non-singular bounce like behaviour, 
and following the bounce, it smoothly enters to an expanding phase. With a suitable conformal transformation of the metric, the F(R) action 
can be mapped to a scalar-tensor theory where the KR field gets coupled with the scalaron (coming from the higher curvature d.o.f) by a simple 
power law coupling. The interaction energy density between the KR and the scalaron seems to grow 
faster than $a^{-6}$ during the contraction phase of the universe. Thereby such interaction energy density has negligible effects at the distant past, 
however it gradually grows as the universe continues to contract and plays a significant role in violating the null energy condition 
or to trigger a non-singular bounce at a certain value of the scale factor. We examine the energy condition which indeed is violated around the bounce, 
however is restored after the ``bouncing regime'' (defined by the era when the energy condition is violated) 
and continues to be fine till the late stage of the universe. Moreover the existence of the ekpyrotic contraction phase ensures the dilution 
of the anisotropic energy density compared to that of the bouncing agent, which in turn leads to a natural bounce scenario that is free from the BKL 
instability. Being the bounce is symmetric and ekpyrotic in nature, the energy density of the bouncing agent rapidly decreases 
(faster than that of the pressureless matter and radiation) after the bounce during the expanding phase, 
and consequently the standard Big-Bang cosmology of the universe gets recovered. 
The scalaron potential is determined which possesses a stable minimum, and due to the ekpyrotic character of the bounce, the scalaron 
potential acquires negative values at the distant universe far away from the bounce. In regard to the perturbation analysis, we find that the comoving 
Hubble radius diverges at the distant past, which indicates that the primordial perturbation modes generate at the deep contracting phase when all 
the perturbation modes lie within the sub-Hubble domain in the Bunch-Davies state. Since the model involves two fields (the KR and the scalaron), 
beside the curvature perturbation, isocurvature perturbation also arises. However the KR field energy density is suppressed compared to the scalaron self 
energy density at the late contracting stage of the universe's evolution when the perturbation modes generate. Thereby we safely ignore the coupling 
between the curvature and the isocurvature perturbation and solely concentrate on the curvature perturbation. It turns out that the curvature perturbation 
gets a blue tilted spectrum over the large scale modes -- not consistent the Planck 2018 data. 
Such inconsistency arises due to the fact that the large scale modes cross the horizon during the $ekpyrotic$ contraction phase when the effective EoS 
parameter is larger than unity. To circumvent this problem, we propose an extended scenario where the ekpyrotic era is preceded by a quasi-matter 
dominated pre-ekpyrotic stage. The transition from the pre-ekpyrotic to the ekpyrotic phase is ensured to be a $smooth$ transition 
by demanding the continuity of the scale factor and the Hubble parameter at the junction time. In such extended scenario, the primordial curvature 
perturbation, on scales that cross the horizon during the quasi-matter dominated pre-ekpyrotic stage, gets a slight red tilted spectrum -- 
this turns out to be consistent with the recent Planck data for suitable parameter values. In particular, if the scale factor of the pre-ekpyrotic 
era is described by $a(\eta) \sim \eta^{2m/\left(1-2m\right)}$, then the parametric regime that makes the model consistent with the Planck data is given by:
$0.3341 \lesssim m \lesssim 0.3344$.

Thus as a whole, the present ``Kalb-Ramond coupled scalaron'' theory predicts a non-singular bounce, in which -- (1) the existence of an 
ekpyrotic phase of contraction resolves the BKL instability, (2) the primordial curvature perturbation gets a nearly scale invariant power spectrum to be 
consistent with the Planck 2018 data and (3) the standard Big-Bang cosmology can be recovered during the expansion of the universe. The work 
clearly demonstrates that the rank two Kalb-Ramond field which has negligible footprints at the present universe, plays the most important role during the 
early stage of the universe to trigger an ekpyrotic bounce, which in turn provides a possible explanation for the removal of the Big-Bang singularity.

\appendix*
\section*{Appendix-I: Solution for the KR field equation}
\label{sec-appendix-1}
The field equation for Kalb-Ramond field is given by,
\begin{eqnarray}
 \partial_{\mu}\left[a^3(t)\Phi(t)H^{\mu\nu\lambda}\right] = 0\ ,
 \label{app2 1}
\end{eqnarray}
Here the greek indices $\nu$, $\lambda$ run from $0$ to $3$. 
\begin{itemize}
 \item For $\nu = 2$ and $\lambda = 3$, Eq.(\ref{app2 1}) becomes
 \begin{eqnarray}
  \partial_{t}\left[a^3(t)\Phi(t)H^{023}\right] + \partial_{x}\left[a^3(t)\Phi(t)H^{123}\right] + 
  \partial_{y}\left[a^3(t)\Phi(t)H^{223}\right] + \partial_{z}\left[a^3(t)\Phi(t)H^{323}\right] = 0\ .
 \label{app2 3}
 \end{eqnarray}
Due to the antisymmetric nature of the KR field, the last two terms of the above equation identically vanish. Furthermore, 
from eqn.~(\ref{solution-off-diagonal-E-equation}), $H^{023} = 0$. As a result, only the second term of Eq.(\ref{app2 3}) 
survives and leads to the information that the non-zero 
component of KR field ($H^{123}$) is independent of the coordinate $x$ i.e $\partial_{x}\left[H^{123}\right] = 0$.

\item For $\nu = 1$ and $\lambda = 3$, Eq.(\ref{app2 1}) becomes
\begin{eqnarray}
  \partial_{t}\left[a^3(t)\Phi(t)H^{013}\right] + \partial_{x}\left[a^3(t)\Phi(t)H^{113}\right] + 
  \partial_{y}\left[a^3(t)\Phi(t)H^{213}\right] + \partial_{z}\left[a^3(t)\Phi(t)H^{313}\right] = 0\ .
 \label{app2 4}
 \end{eqnarray}
 Here the third term survives, which ensures that $H^{123}$ is independent of $y$.
 
 \item For $\nu = 1$ and $\lambda = 2$, Eq.(\ref{app2 1}) becomes
 \begin{eqnarray}
  \partial_{t}\left[a^3(t)\Phi(t)H^{012}\right] + \partial_{x}\left[a^3(t)\Phi(t)H^{112}\right] + 
  \partial_{y}\left[a^3(t)\Phi(t)H^{212}\right] + \partial_{z}\left[a^3(t)\Phi(t)H^{312}\right] = 0\ ,
 \label{app2 5}
 \end{eqnarray}
 where the fourth term sustains and gives $\partial_{z}\left[H^{123}\right] = 0$.
 
\end{itemize}

Therefore it is clear that the non-zero component of the Kalb-Ramond field i.e $H^{123} = h^4$ (or $H_{123} = h_4$) 
depends only on the time ($t$) coordinate.

\section*{Appendix-II: Detailed calculations of Eq.(\ref{KR field equation-2})}\label{sec-appendix-new}
Differentiating both sides (with respect to $t$) of the temporal component of Einstein equation, one gets
\begin{eqnarray}
 6H\dot{H} = \kappa^2\left[\frac{3}{2\kappa^2}\left\{\frac{\dot{\Phi}\ddot{\Phi}}{\Phi^2} - \frac{\dot{\Phi}^3}{\Phi^3}\right\} 
 + V'(\Phi)\dot{\Phi} + \frac{\kappa}{2}\sqrt{\frac{2}{3}}\frac{d}{dt}\left(\Phi h_4h^{4}\right)\right]~.\nonumber
\end{eqnarray}
Using the spatial component of Einstein equation into the above expression, one gets,
\begin{eqnarray}
 3H\left[-\frac{3}{2}\left(\frac{\dot{\Phi}}{\Phi}\right)^2 - \sqrt{\frac{2}{3}}~\kappa^3\Phi~h_4h^4\right] = 
 \frac{3}{2}\left[\frac{\dot{\Phi}}{\Phi}\left(\frac{\ddot{\Phi}}{\Phi} - \frac{\dot{\Phi}^2}{\Phi^2}\right) 
 + \frac{2\kappa^2}{3}~V'(\Phi)\dot{\Phi} + \frac{\sqrt{2}}{3\sqrt{3}}~\kappa^3\frac{d}{dt}\left(\Phi h_4h^4\right)\right]~.\nonumber
\end{eqnarray}
Using the scalar field equation into the above expression,
\begin{eqnarray}
 H\left[-\frac{3}{2}\left(\frac{\dot{\Phi}}{\Phi}\right)^2 - \sqrt{\frac{2}{3}}~\kappa^3\Phi~h_4h^4\right] = 
 \frac{1}{2}\left[\frac{\dot{\Phi}}{\Phi}\left(-3H\frac{\dot{\Phi}}{\Phi} - \frac{\sqrt{2}}{3\sqrt{3}}~\kappa^3\Phi h_4h^4\right) 
 + \frac{\sqrt{2}}{3\sqrt{3}}~\kappa^3\frac{d}{dt}\left(\Phi h_4h^4\right)\right]~~.\nonumber
\end{eqnarray}
Simplifying the above expression, one obtains,
\begin{eqnarray}
 \frac{d}{dt}\left(h_4h^4\right) + 6Hh_4h^4 = 0~~,
\end{eqnarray}
which is shown in Eq.(\ref{KR field equation-2}).

\section*{Appendix-III: With the condition \underline{$3n^2-2n-12 > 0$}}\label{sec-appendix-2}
\begin{figure}[!h]
\begin{center}
 \centering
 \includegraphics[width=3.5in,height=2.5in]{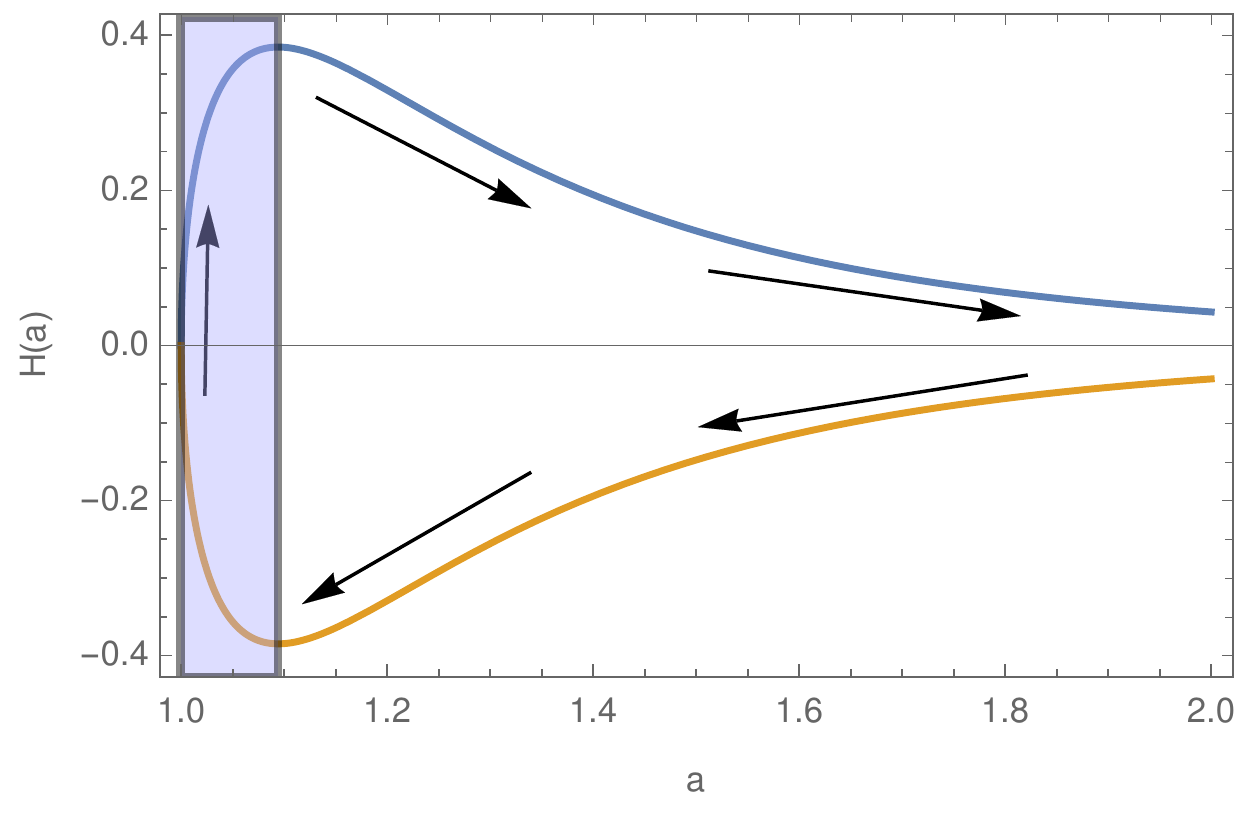}
 \caption{$H(a)$ vs. $a$ from Eq.(\ref{solution-Hubble parameter-case-2}), where we take $a_0 = 1$. 
 The yellow and blue curves represent the negative and the positive branch of $H(a)$ respectively.}
 \label{plot-Hubble-case-2}
\end{center}
\end{figure}

Here we consider $3n^2 - 2n - 12 = 2s$, with $s > 0$. The solution of $H(a)$ from Eq.(\ref{solution-Hubble parameter}) turns out to be,
\begin{eqnarray}
 H(a) = \pm \sqrt{\frac{\kappa^2h_\mathrm{0}}{s}}~a^\mathrm{-3n^2/4}\left\{a^s - a_0^s\right\}^{1/2}
 \label{solution-Hubble parameter-case-2}
\end{eqnarray}
where the integration constant $C_1$ is replaced by $C_\mathrm{1} = \kappa^2h_\mathrm{0}/\left(sa_\mathrm{0}^s\right)$. Clearly 
the Hubble parameter satisfies the following conditions:
\begin{eqnarray}
 H(a=a_0) = 0~~~~~~~~~~\mathrm{and}~~~~~~~~~~~\frac{dH}{dt}\bigg|_{a_0} = \left(aH\right)\frac{dH}{da}\bigg|_{a_0} = \frac{\kappa^2h_0}{2a^{6+n}}~,
 \label{bounce-confirm-2}
\end{eqnarray}
which results to a bouncing universe at $a=a_0$. Thus similar to the earlier case (see Sec.[\ref{sec-background}]), the Hubble parameter 
starts from the negative branch solution at the distant past and leads to a bounce scenario at $a=a_0$. Consequently, $H(a)$ enters to the 
positive branch solution, which refers to an expanding phase of the universe. In particular, the Hubble parameter flows along the directed arrow 
shown in the Fig.[\ref{plot-Hubble-case-2}] (where we take $a_0 = 1$) from $t\rightarrow -\infty$ to $t\rightarrow +\infty$.\\

Actually the null energy condition is violated around $a=a_0$, which is reflected 
through the expression of $\dot{H}$ in Eq.(\ref{bounce-confirm-2}), that makes the bounce possible. 
The presence of $h_0$ in the expression of $\dot{H}(a_0)$ reveals the importance of the KR field in violating the null energy condition or 
equivalently to get the non-singular bounce. To examine the energy condition, we determine $\rho+p$, 
\begin{eqnarray}
 \rho + p = h_0~a^{-3n^2/2}\left\{a^s\left(\frac{3n^2}{2s}-1\right) - a_0^s\left(\frac{3n^2}{2s}\right)\right\}~.
 \label{energy condition-case-2}
\end{eqnarray}
Eq.(\ref{energy condition-case-2}) depicts that 
$\rho+p < 0$ around the bounce, however experiences a zero crossing from negative to positive values at $a=a_c$, where
\begin{eqnarray}
 a_c = a_0\left[\frac{3n^2}{2(n+6)}\right]^{1/s}~,
\end{eqnarray}
and continues to be positive with the evolution of the universe. Thereby the regime where the null energy condition is violated, 
known as the ``bouncing regime'', is given by $a < a_c$ (the shaded region of Fig.[\ref{plot-Hubble-case-2}]).

The Hubble parameter at late contracting stage (i.e $a \gg a_0$) from Eq.(\ref{solution-Hubble parameter-case-2}) becomes,
\begin{eqnarray}
 H(a) \propto \pm a^{-(6+n)/2}~,
 \label{late-Hubble parameter-case-2}
\end{eqnarray}
which immediately leads to the effective EoS parameter during the contracting phase of the universe as,
\begin{eqnarray}
 \omega_\mathrm{eff} = -1 - \left(\frac{2a}{3H}\right)\frac{dH}{da} = 1 + \frac{n}{3}~.
 \label{eos-case-1}
\end{eqnarray}
Clearly $\omega_\mathrm{eff} > 1$, or equivalently the bounce is ekpyrotic in nature, for $n > 0$. Moreover the condition $3n^2-2n-12>0$ is reduced to 
$n > \frac{1}{3}\left(1 + \sqrt{37}\right)$. This is the range of $n$ which leads to an ekpyrotic character of the bounce and helps to resolve the 
BKL instability in the present case. From Eq.(\ref{solution-Hubble parameter-case-2}), we get the energy density of the bouncing agent 
scales by $a^{-(6+n)}$ after the bounce during the expanding phase of the universe. Therefore we may argue that the total 
energy density of the bouncing agent 
decreases at a faster rate compared to that of the pressureless matter and radiation during the expansion of the universe, and hence, the standard Big-Bang 
cosmology gets recovered.

\end{document}